# Sampling by Reversing The Landmarking Process

Lee, C.K.
chengli@uab.edu

**Abstract**

Variations of the commonly applied landmark sampling are presented. These samplings are "forward" landmarking such that each stack of data is created by first selecting landmarks and then including the subsequent observations of the selected landmarks. Unlike these forward landmarking samplings, a "backward" or "reverse" landmarking is proposed with a flexible "progressive" weighting on selecting different types of events and non-events. The backward landmark sample is compared with those forward landmark samples with a real world mortgage data. Results show that the backward landmark sample has smaller sampling errors than of those forward landmark samples.

## I. Introduction

Landmark sampling was introduced by van Houwelinge and Putter (2011). It has been widely applied in the medical field which interest is often in the probability of the occurrence of an event in a prediction window following a per-determined time point. That is, the landmark time point (the beginning of the time) of each stack of the data is treated the same when all the stacks are piled up to form the so-called supermodel or super dataset. The assumption is that the distribution of a medical event (occurrence of cancer, disease, death, etc.) is the same after a landmark time point that can be at any time point during a span of years. In other words, a landmark time point can be at any month and the seasonality effect is ignored. For example, the distribution of an event after, say, January 2000 is the same as the distribution after, say, March 2010.

In the financial data, however, the distributions of the occurrence of an event may be different after different time points. For example, the number of house foreclosures jumped during the financial crisis in 2008. Therefore, when applying landmark sampling in financial data, the landmark time points (landmarks, in short) should be treated differently such that the seasonality effects remain intact. This study focuses only on the various landmark samplings in which the event rates are of the interest. No modeling methodology is recommended.

The layout of this paper is as follows. Section II provides an illustration on how to align all the stacks to form a super dataset such that the seasonality is preserved. The illustration is demonstrated with a real word dataset from Freddie Mac in Section 3. In Section 4, the backward landmark sampling with a flexible weighting scheme is proposed for selecting a landmark sample when the full supermodel (the super dataset) cannot be practically constructed.

## II. An Illustration

Super dataset is defined as the compilation of the stacked datasets from all landmark time points. The construction is illustrated using a made-up example of 4 loans shown in Table 1.1. Assuming the censoring date is June 2001, the 4 loans are alive in January 2001 with Loan ID 1 experiencing the event of interest (the event, hereafter) in June, Loan ID 2 experiencing the event in April, Loan ID 3 is censored in June, and Loan ID 4 is censored in March due to the occurrence of other events or lost to follow-up. Note that 0/1 in each cell are the values of the event indicator with 1 denoting the occurrence of the event and 0 otherwise.

Since each observation is created at month-end, the total number in each date can be regarded as the number at risk. Therefore, the hazard of each date is the number of events divided by the total number.

When creating all the stacks from each loan beginning at all the landmark dates, the compilation of the super dataset aligned by date is in Table 1.2. Since the super dataset is aligned by date, let's call it the Date Aligned Super Dataset (DASD). Note that the number of events and the total number in DASD at $i$th $t$ is equal to the corresponding number in unstacked original dataset in Table 1.1 multiplying $t$. There are two folds of benefits by creating a DASD. One is that the monthly hazard is identical between the original data and the DASD. The other is that the time-stamped explanatory variables such as macro-economic variables can be merged with DASD by date.

The generalization of the construction of DASD is in Appendix E. Given a dataset beginning at a date and truncated at a later date, from this exercise, a DASD is constructed. The next section is the application on a real word dataset.

Table 1.1

| Date | ID 1 | ID 2 | ID 3 | ID 4 | Number of Events | Total Number | Hazard |
|---|---|---|---|---|---|---|---|
| 01/2001 | 0 | 0 | 0 | 0 | 0 | 4 | 0 |
| 02/2001 | 0 | 0 | 0 | 0 | 0 | 4 | 0 |
| 03/2001 | 0 | 0 | 0 | 0 | 0 | 4 | 0 |
| 04/2001 | 0 | 1 | 0 |   | 1 | 3 | 0.3333 |
| 05/2001 | 0 |   | 0 |   | 0 | 2 | 0 |
| 06/2001 | 1 |   | 0 |   | 1 | 2 | 0.5 |

Table 1.2

| Landmark Date | Stacks of ID 1 | | | | | | Stacks of ID 2 | | | | Stacks of ID 3 | | | | | | Stacks of ID 4 | | | Number of Events | Total Number | Hazard |
|---|---|---|---|---|---|---|---|---|---|---|---|---|---|---|---|---|---|---|---|---|---|---|
|  | Jan | Feb | Mar | Apr | May | Jun | Jan | Feb | Mar | Apr | Jan | Feb | Mar | Apr | May | Jun | Jan | Feb | Mar |  |  |  |
| 1/2001, $t$=1 | 0, 1 |  |  |  |  |  | 0, 1 |  |  |  | 0, 1 |  |  |  |  |  | 0, 1 |  |  | 0 | 4 | 0 |
| 2/2001, $t$=2 | 0, 2 | 0, 1 |  |  |  |  | 0, 2 | 0, 1 |  |  | 0, 2 | 0, 1 |  |  |  |  | 0, 2 | 0, 1 |  | 0 | 8 | 0 |
| 3/2001, $t$=3 | 0, 3 | 0, 2 | 0, 1 |  |  |  | 0, 3 | 0, 2 | 0, 1 |  | 0, 3 | 0, 2 | 0, 1 |  |  |  | 0, 3 | 0, 2 | 0, 1 | 0 | 12 | 0 |
| 4/2001, $t$=4 | 0, 4 | 0, 3 | 0, 2 | 0, 1 |  |  | 1, 4 | 1, 3 | 1, 2 | 1, 1 | 0, 4 | 0, 3 | 0, 2 | 0, 1 |  |  |  |  |  | 4 | 12 | 0.3333 |
| 5/2001, $t$=5 | 0, 5 | 0, 4 | 0, 3 | 0, 2 | 0, 1 |  |  |  |  |  | 0, 5 | 0, 4 | 0, 3 | 0, 2 | 0, 1 |  |  |  |  | 0 | 10 | 0 |
| 6/2001, $t$=6 | 1, 6 | 1, 5 | 1, 4 | 1, 3 | 1, 2 | 1, 1 |  |  |  |  | 0, 6 | 0, 5 | 0, 4 | 0, 3 | 0, 2 | 0, 1 |  |  |  | 6 | 12 | 0.5 |

### III. Application

Original dataset and Super dataset

The dataset of harp_historical_data1.txt was downloaded from Freddie Mac's website with monthly performance information of mortgage from June 2012 to June 2018. After cleaning up the data, loans beginning with loan age 0 (that is, when they were booked) and with continuous monthly performance records are kept. The total number of loans (identified as ID_Loan in the dataset) is 959,959, and the total number of their monthly observations is 56,119,398. Among these 959,959 loans, during the data collection period, there are 4 major events REO (real estate owned, 0.832%), Chargeoff (0.891%), Payoff (41.183%), Others (0.214%), and survival (56.880%). Events are recorded at the end of the loan's lifetime. The numbers of the 4 types of events and survivals by date are in Appendix A.

From the original dataset, a landmark dataset can be created by selecting the individuals from a time point known as landmark, and, then selecting all their subsequent observations up to the event time or censoring time. This process is call landmarking (van Houwelinge and Putter, 2011). When selecting all the landmarks and stacking up all the landmark datasets, it becomes the super dataset (super prediction dataset or super stacked dataset by some researchers). Following the illustration in Section I, the super dataset created from the original dataset has the total number of observations 1,898,799,764. Given a landmark and a prediction window, as it has been shown in the illustration table in Section II, numbers of events and survivals of the super dataset are the numbers in the unstacked original dataset multiplying *ith* month denoted as *t*. An example with the landmark as of July 2016 is shown in Appendix B.

Ideally, analyses should be done on the super dataset for all the stacks from all landmarks. However, it is impractical to analyze such a large dataset, and in some cases it is infeasible to create and store a super dataset; therefore, it is desirable to create samples from the super dataset. In the next section, five different samplings are presented.

### IV. Samplings

Five samplings are discussed in this section. Each sampling creates a date aligned stack dataset.

From the unstacked original dataset of 56,119,398 observations, the super dataset is created by including all the landmarks and all the observations following the landmarks. Note that the observations at each landmark are survivals only. The super dataset creation is mentioned by Liu (et al., 2019) by allowing the prediction window to vary and not necessarily fall within a fixed time suggested by van Houwelingen and Putter (2011). The prediction window after a landmark is referred as "month after landmark". For convenience purpose, month after landmark begins at 1 at the landmark.

As shown in Section III, the super dataset is large and it is impractical for analyzing the entire data. Similar to how the super dataset is created, therefore, a sample of landmarks can be selected first, and then, all the subsequent observations are selected. Recall that super dataset includes all the landmarks. For identifiability purpose, the data are sorted by loan ID and date before the samples of landmarks are selected. Five commonly applied stacked samples are presented. These five samples are created by first selecting a number of landmarks, and, then, all the observations after each landmark are appended to form the stacked samples. The difference is how the landmarks are selected.

Sampling 1 – Uniformly Spaced Landmark Sample (uniform sample, hereafter)

Landmarks are selected with an equal distance between any two landmarks. For example, if the first selected landmark is January and the distance is set at 3, then, April, July, October are the selected landmarks. When the distance is 1, then, it becomes the super dataset. Under this sampling, a uniform sample is created by selecting the first time point from each ID as the first landmark, and then the following landmarks are selected with a distance of 6. The uniform sampling is suggested by van Houwelingen and Putter (2011).

Sampling 2 – Vertical Random Landmark Sample (vertical sample, hereafter)

Unlike uniformly spaced sampling, landmarks are randomly selected from all possible time points. When all the landmarks are selected, it becomes the super dataset. Under this sampling, a 20% random landmarks are selected to create a vertical sample.

Sampling 3 – Horizontal Random Landmark Sample (horizontal sample, hereafter)

Landmarks are randomly selected from each prediction time point (month, for example). For example, if a dataset beings in January 2000, then, a set of random landmarks are selected from January 2000, February 2000, and so on. This sampling can be regarded as the stratified sampling that each month is a stratum. When all the landmarks are selected, it also becomes the super dataset. Under this sampling, a 20% random landmarks are selected from each time point to form a horizontal sample.

Sampling 4 – Single Random Landmark Weighted Sample (single sample, hereafter)

One landmark is selected from each ID (subject, loan, individual, person, etc.). Then, all the following observations are appended to the selected landmark to form a stacked sample. Each observation within the same ID is then weighted by the number of the observations of the ID. Such a weighting is a common practice in regression modeling (Lohr, 2009). Contrary to super dataset, only one landmark from each ID is selected in this sampling, and, therefore, it is a subset of the super dataset. Note that this sampling can be regarded as the stratified sampling where observations within each ID is a stratum. Under this sampling, a single sample is created.

Sampling 5 – Backward Landmark Sample (backward sample, hereafter)

The four samplings mentioned above are a subset of the super dataset and the process is forward landmarking meaning the landmarks are selected months forward. Unlike these forward landmarking samples, the backward landmark sampling is proposed by first select events and non-events from each date and, then, landmarks are selected backward from those dates of events and non-events.

Based on the table in Appendix A (regarded as the size table), the numbers of events vary across time and are much smaller than of the survivals. For illustration purpose, the following progressive weighting is to deliberately apply more weights on events with relatively less occurrence.

Table 4 Progressive Weighting

| Number of Events or Survivals | % Random Events or Survivals Selected | Weight Assigned |
|---|---|---|
| [1, 100] | 100 | 1/1 |
| [101, 500] | 90 | 1/0.9 |
| [501, 1000] | 80 | 1/0.8 |
| [1001, 2000] | 70 | 1/0.7 |
| [2001, 3000] | 60 | 1/0.6 |
| [3001, 4000] | 50 | 1/0.5 |
| [4001, 5000] | 40 | 1/0.4 |
| [5001, 6000] | 30 | 1/0.3 |
| [6001, 7000] | 20 | 1/0.2 |
| 7001 and up | 10 | 1/0.1 |

Applying the progressive weights on the events and survivals, the events are over-sampled while the survivals are under-sampled. Note that the weight is the reciprocal of the percentage selected. Obviously, the backward sample is also a subset of the super dataset.

The sizes of the samples and the comparisons of monthly hazard rates are in the next Section.

## V. Comparisons of The Five Samplings

Table 5.1 Numbers/weights of unstacked, four forward samples, and the backward sample

|  | # Observations | Total Weight |
|---|---|---|
| Original | 56,119,398 | N/A |
| Unform | 340,315,507 | N/A |
| Vertical | 379,716,715 | N/A |
| Horizontal | 379,765,804 | N/A |
| Single Sample | 28,556,220 | 1,899,209,658 |
| Backward Sample | 295,825,769 | 1,898,888,910 |
| Super Dataset | 1,898,788,764 | N/A |

Table 5.1 shows the number of observations of the super dataset is almost 1.9 billions. It is impractical to analyze such a huge data set. Therefore, the 5 stacked samples are the candidates. Among the 5 stacked samples, the total weight of the single sample and the backward sample is close to the total observations of the super dataset by design.

Two landmarks are selected for the comparisons. One is January 2010 and the other is January 2013. The prediction time horizon for both is 24 months. The monthly hazard rate for each event, and the event numbers or weights (for single and backward sample) are in Appendix C.

As it is mentioned previously, the super dataset is the compilation of all the stacked samples from all the landmarks from the unstacked original data. And, all the five samples are the sub-sample of the super dataset. Given a landmark and a prediction time horizon, the sampling errors are measured by MAE (mean absolute error) and RMSE (root mean square error) between a sample and the unstacked original dataset.

From the DASD of the 5 samples shown in Appendix C, MAEs and RMSEs are provided in Table 5.2. MAEs and RMSEs are comparable among uniform, vertical and horizontal sample. Single sample has the largest MAE and RMSE while backward sample has the smallest MAE and RMSE as expected. The curves of the hazards of the events are in Appendix D. Although they are not presented in this study, the standard errors of hazards are provided in Appendix E for constructing the upper and lower limits. Appendix E also shows the hazard estimators of the forward samplings are unbiased to of the original dataset.

Table 5.2 MAE and RMSE of Two Landmarks

| Event | Sample | January 2010 | | January 2013 | |
|---|---|---|---|---|---|
| | | MAE | RMSE | MAE | RMSE |
| REO | Uniform | 0.000102 | 0.00014 | 0.000029 | 0.000034 |
| | Vertical | 0.000105 | 0.000141 | 0.000022 | 0.000031 |
| | Horizontal | 0.000108 | 0.000149 | 0.000028 | 0.000035 |
| | Single | 0.000253 | 0.000334 | 0.000097 | 0.000124 |
| | Backward | 6.00E-07 | 6.84E-07 | 2.78E-06 | 4.47E-06 |
| Chargeoff | Uniform | 0.000082 | 0.0001 | 0.000028 | 0.000035 |
| | Vertical | 0.000086 | 0.000129 | 0.000028 | 0.000037 |
| | Horizontal | 0.000126 | 0.000185 | 0.000031 | 0.000041 |
| | Single | 0.000237 | 0.000356 | 0.000083 | 0.00011 |
| | Backward | 7.02E-06 | 0.00002 | 3.05E-06 | 4.18E-06 |
| Payoff | Uniform | 0.000067 | 0.000093 | 0.000022 | 0.00003 |
| | Vertical | 0.00008 | 0.000114 | 0.000021 | 0.00003 |
| | Horizontal | 0.00008 | 0.000107 | 0.000019 | 0.000024 |
| | Single | 0.000281 | 0.000524 | 0.000064 | 0.000102 |
| | Backward | 1.97E-06 | 9.68E-06 | 1.59E-06 | 3.30E-06 |
| Others | Uniform | 0.000445 | 0.000516 | 0.000328 | 0.000428 |
| | Vertical | 0.000563 | 0.000773 | 0.000189 | 0.000253 |
| | Horizontal | 0.000487 | 0.000668 | 0.000161 | 0.000217 |
| | Single | 0.001189 | 0.001764 | 0.00051 | 0.000704 |
| | Backward | 0.000147 | 0.000219 | 0.000138 | 0.000201 |

**Discussion and Conclusion**

In single sample, each individual can be regarded as a stratum and applying a weight equal to the number of the observations within the stratum such that the sum of the sampling weights is equal to the population size. Because one observation is selected from each individual, every individual of event or non-event has its representation in the sample. However, if an individual (stratum) has 20 observations and the first observation is randomly selected, the sum of sampling weights for the loan is 400 while the sum is 20 if the last observation is selected. Therefore, the variation is large and that is why the sampling errors of the single sample are the largest.

Among the uniform sample, the vertical sample and the horizontal sample, the uniform sample has the smallest MAE and RMSE in most cases. However, In the uniform sample, the choice of the distance between two landmarks is arbitrary. And, the seasonality effect may be neglected when the distance excludes certain months or quarters. For the data associated with date, the uniform sample may not be a good choice.

Then, the vertical sampling and the horizontal sampling are the better choice for creating a landmark sample. Under the horizontal sampling, each time point has a random sample of landmarks being selected; therefore, the seasonality should be well preserved. As for the vertical sampling, it is simple to implement, and it should as well preserve the seasonality. The drawback of the vertical and the horizontal sampling is that the landmarks of a short lifetime individual is less likely to be selected comparing to a long lifetime individual. The likelihood of the landmarks being selected becomes even less when individuals have a short lifetime and their events are considered as a rare event.

In this study, the proposed backward sampling has the least sampling errors and the reason is that the numbers of events are first selected so that they are not underrepresented. The sample size is even less than those of the vertical and horizontal sample. Although the backward sampling requires the creating of the size table shown in Appendix A, this extra step is worth it for the small sampling errors.

# Appendix A – Number of Events by Date of The Unstacked Original Freddie Mac Data

| Date | REO | Chargeoff | Payoff | Others | Survival | Total |
|---|---|---|---|---|---|---|
| Apr-2009 | 0 | 0 | 0 | 0 | 154 | 154 |
| May-2009 | 0 | 0 | 0 | 0 | 2,829 | 2,829 |
| Jun-2009 | 0 | 0 | 1 | 0 | 9,607 | 9,608 |
| Jul-2009 | 0 | 0 | 3 | 0 | 20,438 | 20,441 |
| Aug-2009 | 0 | 0 | 11 | 1 | 30,184 | 30,196 |
| Sep-2009 | 0 | 0 | 19 | 1 | 36,217 | 36,237 |
| Oct-2009 | 0 | 0 | 27 | 4 | 41,093 | 41,124 |
| Nov-2009 | 0 | 1 | 37 | 4 | 47,916 | 47,958 |
| Dec-2009 | 0 | 1 | 46 | 4 | 56,850 | 56,901 |
| Jan-2010 | 1 | 4 | 42 | 2 | 68,060 | 68,109 |
| Feb-2010 | 1 | 2 | 35 | 1 | 78,387 | 78,426 |
| Mar-2010 | 3 | 4 | 90 | 0 | 86,585 | 86,682 |
| Apr-2010 | 5 | 2 | 120 | 3 | 96,713 | 96,843 |
| May-2010 | 3 | 5 | 162 | 1 | 105,871 | 106,042 |
| Jun-2010 | 3 | 3 | 193 | 7 | 112,512 | 112,718 |
| Jul-2010 | 13 | 4 | 217 | 1 | 121,638 | 121,873 |
| Aug-2010 | 12 | 7 | 345 | 7 | 133,211 | 133,582 |
| Sep-2010 | 12 | 9 | 407 | 18 | 146,572 | 147,018 |
| Oct-2010 | 23 | 10 | 494 | 14 | 160,961 | 161,502 |
| Nov-2010 | 10 | 14 | 623 | 12 | 176,046 | 176,705 |
| Dec-2010 | 24 | 17 | 615 | 9 | 193,851 | 194,516 |
| Jan-2011 | 23 | 22 | 429 | 57 | 210,607 | 211,138 |
| Feb-2011 | 17 | 17 | 328 | 26 | 225,827 | 226,215 |
| Mar-2011 | 27 | 19 | 348 | 25 | 237,676 | 238,095 |
| Apr-2011 | 32 | 25 | 375 | 29 | 246,358 | 246,819 |
| May-2011 | 37 | 26 | 459 | 34 | 253,460 | 254,016 |
| Jun-2011 | 37 | 38 | 585 | 55 | 259,404 | 260,119 |
| Jul-2011 | 48 | 46 | 610 | 44 | 266,808 | 267,556 |
| Aug-2011 | 76 | 49 | 846 | 78 | 273,897 | 274,946 |
| Sep-2011 | 56 | 49 | 956 | 44 | 282,334 | 283,439 |
| Oct-2011 | 40 | 55 | 1,237 | 39 | 292,230 | 293,601 |
| Nov-2011 | 62 | 64 | 1,294 | 61 | 302,516 | 303,997 |
| Dec-2011 | 53 | 55 | 1,256 | 76 | 313,359 | 314,799 |
| Jan-2012 | 71 | 61 | 1,136 | 58 | 325,127 | 326,453 |
| Feb-2012 | 62 | 55 | 1,409 | 78 | 339,221 | 340,825 |
| Mar-2012 | 81 | 78 | 1,641 | 71 | 361,408 | 363,279 |
| Apr-2012 | 52 | 86 | 1,661 | 64 | 383,483 | 385,346 |
| May-2012 | 67 | 83 | 1,874 | 446 | 409,497 | 411,967 |
| Jun-2012 | 75 | 109 | 2,022 | 57 | 441,149 | 443,412 |
| Jul-2012 | 67 | 110 | 2,338 | 76 | 470,410 | 473,001 |
| Aug-2012 | 90 | 129 | 2,753 | 79 | 494,531 | 497,582 |
| Sep-2012 | 72 | 137 | 2,715 | 41 | 519,986 | 522,951 |
| Oct-2012 | 74 | 122 | 3,299 | 43 | 540,723 | 544,261 |
| Nov-2012 | 82 | 142 | 3,197 | 40 | 566,822 | 570,283 |
| Dec-2012 | 60 | 191 | 3,478 | 34 | 587,359 | 591,122 |
| Jan-2013 | 84 | 133 | 3,483 | 38 | 608,204 | 611,942 |
| Feb-2013 | 83 | 143 | 3,350 | 34 | 631,317 | 634,927 |
| Mar-2013 | 97 | 159 | 3,861 | 37 | 654,250 | 658,404 |
| Apr-2013 | 103 | 177 | 4,365 | 34 | 676,907 | 681,586 |
| May-2013 | 75 | 194 | 5,047 | 16 | 697,160 | 702,492 |
| Jun-2013 | 104 | 191 | 5,030 | 31 | 715,121 | 720,477 |
| Jul-2013 | 124 | 211 | 5,409 | 30 | 728,877 | 734,651 |
| Aug-2013 | 136 | 172 | 4,274 | 23 | 741,568 | 746,173 |
| Sep-2013 | 113 | 155 | 3,263 | 27 | 752,751 | 756,309 |
| Oct-2013 | 138 | 166 | 3,235 | 25 | 759,076 | 762,640 |
| Nov-2013 | 126 | 123 | 3,038 | 24 | 764,327 | 767,638 |
| Dec-2013 | 136 | 157 | 3,288 | 26 | 768,077 | 771,684 |
| Jan-2014 | 136 | 112 | 2,308 | 45 | 772,572 | 775,173 |
| Feb-2014 | 110 | 102 | 2,210 | 53 | 776,287 | 778,762 |
| Mar-2014 | 124 | 106 | 3,055 | 200 | 777,735 | 781,220 |
| Apr-2014 | 102 | 123 | 3,593 | 61 | 779,037 | 782,916 |
| May-2014 | 102 | 126 | 4,033 | 53 | 779,190 | 783,504 |
| Jun-2014 | 83 | 109 | 4,741 | 59 | 778,144 | 783,136 |
| Jul-2014 | 76 | 119 | 5,345 | 78 | 775,803 | 781,421 |
| Aug-2014 | 102 | 113 | 4,954 | 76 | 773,413 | 778,658 |
| Sep-2014 | 93 | 119 | 4,731 | 75 | 771,562 | 776,580 |
| Oct-2014 | 119 | 137 | 4,917 | 85 | 769,508 | 774,766 |
| Nov-2014 | 85 | 117 | 4,201 | 61 | 768,229 | 772,693 |
| Dec-2014 | 141 | 136 | 4,936 | 80 | 765,616 | 770,909 |
| Jan-2015 | 120 | 90 | 3,847 | 84 | 764,257 | 768,398 |
| Feb-2015 | 102 | 114 | 4,555 | 93 | 761,985 | 766,849 |
| Mar-2015 | 110 | 125 | 6,743 | 103 | 757,569 | 764,650 |
| Apr-2015 | 102 | 130 | 6,555 | 112 | 753,651 | 760,550 |
| May-2015 | 97 | 106 | 6,809 | 278 | 749,190 | 756,480 |
| Jun-2015 | 98 | 127 | 7,639 | 105 | 743,427 | 751,396 |
| Jul-2015 | 110 | 130 | 7,247 | 126 | 738,007 | 745,620 |
| Aug-2015 | 104 | 114 | 6,489 | 102 | 733,129 | 739,938 |
| Sep-2015 | 99 | 98 | 6,299 | 154 | 728,412 | 735,062 |
| Oct-2015 | 108 | 108 | 6,000 | 164 | 723,826 | 730,206 |
| Nov-2015 | 100 | 67 | 4,879 | 185 | 720,335 | 725,566 |
| Dec-2015 | 85 | 134 | 5,955 | 125 | 715,233 | 721,532 |
| Jan-2016 | 108 | 90 | 4,132 | 106 | 712,180 | 716,616 |
| Feb-2016 | 94 | 107 | 4,474 | 316 | 708,437 | 713,428 |
| Mar-2016 | 118 | 121 | 6,151 | 94 | 703,430 | 709,914 |
| Apr-2016 | 98 | 99 | 6,572 | 133 | 697,947 | 704,849 |
| May-2016 | 100 | 99 | 7,354 | 140 | 691,442 | 699,135 |
| Jun-2016 | 93 | 110 | 8,231 | 77 | 684,046 | 692,557 |
| Jul-2016 | 94 | 80 | 7,605 | 84 | 677,214 | 685,077 |
| Aug-2016 | 107 | 109 | 8,850 | 124 | 668,917 | 678,107 |
| Sep-2016 | 86 | 82 | 8,318 | 76 | 661,295 | 669,857 |
| Oct-2016 | 95 | 90 | 8,229 | 70 | 653,623 | 662,107 |
| Nov-2016 | 103 | 83 | 8,054 | 69 | 646,142 | 654,451 |
| Dec-2016 | 81 | 65 | 7,770 | 366 | 638,539 | 646,821 |
| Jan-2017 | 100 | 57 | 5,602 | 80 | 633,532 | 639,371 |
| Feb-2017 | 83 | 54 | 4,603 | 68 | 629,452 | 634,260 |
| Mar-2017 | 116 | 70 | 5,597 | 71 | 624,252 | 630,106 |
| Apr-2017 | 75 | 67 | 5,195 | 61 | 619,470 | 624,868 |
| May-2017 | 90 | 59 | 6,363 | 276 | 613,137 | 619,925 |
| Jun-2017 | 74 | 52 | 6,900 | 61 | 606,431 | 613,518 |
| Jul-2017 | 79 | 42 | 6,320 | 67 | 600,265 | 606,773 |
| Aug-2017 | 107 | 49 | 7,054 | 139 | 593,221 | 600,570 |
| Sep-2017 | 82 | 47 | 5,956 | 44 | 587,426 | 593,555 |
| Oct-2017 | 78 | 39 | 6,414 | 166 | 581,037 | 587,734 |
| Nov-2017 | 83 | 49 | 5,665 | 108 | 575,440 | 581,345 |
| Dec-2017 | 53 | 45 | 5,501 | 190 | 569,911 | 575,700 |
| Jan-2018 | 84 | 35 | 4,396 | 67 | 565,560 | 570,142 |
| Feb-2018 | 70 | 36 | 4,065 | 68 | 561,321 | 565,560 |
| Mar-2018 | 83 | 36 | 4,862 | 370 | 555,970 | 561,321 |
| Apr-2018 | 98 | 26 | 4,850 | 54 | 550,942 | 555,970 |
| May-2018 | 84 | 29 | 5,456 | 148 | 545,225 | 550,942 |
| Jun-2018 | 72 | 32 | 5,327 | 363 | 539,431 | 545,225 |

# Appendix B – Super Dataset of Landmark at July 2016

| Date | t | Original Unstacked | | | | | | Super Dataset | | | | | |
|---|---|---|---|---|---|---|---|---|---|---|---|---|---|
| | | REO | Charge Off | Payoff | Others | Survival | Total | REO | Charge Off | Payoff | Others | Survival | Total |
| Jul-2016 | 1 | 0 | 0 | 0 | 0 | 677,214 | 677,214 | 0 | 0 | 0 | 0 | 677,214 | 677,214 |
| Aug-2016 | 2 | 107 | 109 | 8,850 | 124 | 668,024 | 677,214 | 214 | 218 | 17,700 | 248 | 1,336,048 | 1,354,428 |
| Sep-2016 | 3 | 86 | 82 | 8,318 | 76 | 659,462 | 668,024 | 258 | 246 | 24,954 | 228 | 1,978,386 | 2,004,072 |
| Oct-2016 | 4 | 95 | 90 | 8,229 | 70 | 650,978 | 659,462 | 380 | 360 | 32,916 | 280 | 2,603,912 | 2,637,848 |
| Nov-2016 | 5 | 103 | 83 | 8,051 | 69 | 642,672 | 650,978 | 515 | 415 | 40,255 | 345 | 3,213,360 | 3,254,890 |
| Dec-2016 | 6 | 81 | 65 | 7,769 | 366 | 634,391 | 642,672 | 486 | 390 | 46,614 | 2,196 | 3,806,346 | 3,856,032 |
| Jan-2017 | 7 | 100 | 57 | 5,597 | 80 | 628,557 | 634,391 | 700 | 399 | 39,179 | 560 | 4,399,899 | 4,440,737 |
| Feb-2017 | 8 | 83 | 54 | 4,596 | 68 | 623,756 | 628,557 | 664 | 432 | 36,768 | 544 | 4,990,048 | 5,028,456 |
| Mar-2017 | 9 | 116 | 70 | 5,594 | 71 | 617,905 | 623,756 | 1,044 | 630 | 50,346 | 639 | 5,561,145 | 5,613,804 |
| Apr-2017 | 10 | 75 | 67 | 5,183 | 61 | 612,519 | 617,905 | 750 | 670 | 51,830 | 610 | 6,125,190 | 6,179,050 |
| May-2017 | 11 | 90 | 59 | 6,356 | 276 | 605,738 | 612,519 | 990 | 649 | 69,916 | 3,036 | 6,663,118 | 6,737,709 |
| Jun-2017 | 12 | 74 | 52 | 6,884 | 60 | 598,668 | 605,738 | 888 | 624 | 82,608 | 720 | 7,184,016 | 7,268,856 |
| Jul-2017 | 13 | 79 | 42 | 6,301 | 66 | 592,180 | 598,668 | 1,027 | 546 | 81,913 | 858 | 7,698,340 | 7,782,684 |
| Aug-2017 | 14 | 106 | 49 | 7,025 | 137 | 584,861 | 592,178 | 1,484 | 686 | 98,350 | 1,918 | 8,188,054 | 8,290,492 |
| Sep-2017 | 15 | 82 | 47 | 5,933 | 44 | 578,755 | 584,861 | 1,230 | 705 | 88,995 | 660 | 8,681,325 | 8,772,915 |
| Oct-2017 | 16 | 76 | 39 | 6,390 | 166 | 572,084 | 578,755 | 1,216 | 624 | 102,240 | 2,656 | 9,153,344 | 9,260,080 |
| Nov-2017 | 17 | 83 | 48 | 5,646 | 108 | 566,199 | 572,084 | 1,411 | 816 | 95,982 | 1,836 | 9,625,383 | 9,725,428 |
| Dec-2017 | 18 | 51 | 45 | 5,478 | 189 | 560,436 | 566,199 | 918 | 810 | 98,604 | 3,402 | 10,087,848 | 10,191,582 |
| Jan-2018 | 19 | 84 | 34 | 4,379 | 66 | 555,873 | 560,436 | 1,596 | 646 | 83,201 | 1,254 | 10,561,587 | 10,648,284 |
| Feb-2018 | 20 | 70 | 35 | 4,041 | 68 | 551,659 | 555,873 | 1,400 | 700 | 80,820 | 1,360 | 11,033,180 | 11,117,460 |
| Mar-2018 | 21 | 83 | 36 | 4,835 | 370 | 546,335 | 551,659 | 1,743 | 756 | 101,535 | 7,770 | 11,473,035 | 11,584,839 |
| Apr-2018 | 22 | 97 | 26 | 4,814 | 52 | 541,346 | 546,335 | 2,134 | 572 | 105,908 | 1,144 | 11,909,612 | 12,019,370 |
| May-2018 | 23 | 84 | 28 | 5,425 | 146 | 535,663 | 541,346 | 1,932 | 644 | 124,775 | 3,358 | 12,320,249 | 12,450,958 |
| Jun-2018 | 24 | 68 | 32 | 5,287 | 362 | 529,914 | 535,663 | 1,632 | 768 | 126,888 | 8,688 | 12,717,936 | 12,855,912 |

# Appendix C

## Landmark Date – January 2010

| Date | REO | | | | | | Charge Off | | | | | | Payoff | | | | | | Others | | | | | | Survival | | | | | |
|---|---|---|---|---|---|---|---|---|---|---|---|---|---|---|---|---|---|---|---|---|---|---|---|---|---|---|---|---|---|---|
| | Original | Uniform | Vertical | Horizontal | Single | Backward | Original | Uniform | Vertical | Horizontal | Single | Backward | Original | Uniform | Vertical | Horizontal | Single | Backward | Original | Uniform | Vertical | Horizontal | Single | Backward | Original | Uniform | Vertical | Horizontal | Single | Backward |
| Jan-2010 | 0 | 0 | 0 | 0 | 0 | 0 | 0 | 0 | 0 | 0 | 0 | 0 | 0 | 0 | 0 | 0 | 0 | 0 | 0 | 0 | 0 | 0 | 0 | 0 | 68060 | 22040 | 13648 | 13794 | 68416 | 68195 |
| Feb-2010 | 1 | 0 | 0 | 0 | 9 | 1 | 2 | 1 | 1 | 0 | 0 | 2 | 35 | 9 | 6 | 8 | 27 | 35 | 1 | 0 | 0 | 1 | 0 | 1 | 68021 | 22030 | 13641 | 13785 | 68380 | 68156 |
| Mar-2010 | 3 | 2 | 1 | 1 | 9 | 3 | 4 | 2 | 3 | 1 | 1 | 0 | 4 | 85 | 23 | 13 | 16 | 95 | 85 | 0 | 0 | 0 | 0 | 0 | 0 | 67929 | 22002 | 13626 | 13767 | 68276 | 68064 |
| Apr-2010 | 5 | 2 | 2 | 1 | 0 | 5 | 2 | 0 | 0 | 0 | 0 | 2 | 106 | 40 | 21 | 16 | 101 | 106 | 3 | 0 | 1 | 0 | 0 | 3 | 67813 | 21960 | 13604 | 13748 | 68175 | 67947 |
| May-2010 | 3 | 3 | 1 | 1 | 0 | 3 | 5 | 4 | 1 | 2 | 0 | 5 | 129 | 35 | 23 | 25 | 164 | 134 | 1 | 0 | 0 | 1 | 0 | 1 | 67675 | 21918 | 13577 | 13721 | 68011 | 67804 |
| Jun-2010 | 3 | 1 | 0 | 2 | 0 | 3 | 3 | 2 | 0 | 0 | 0 | 3 | 155 | 39 | 36 | 30 | 127 | 157 | 7 | 3 | 0 | 4 | 30 | 7 | 67507 | 21873 | 13535 | 13691 | 67854 | 67633 |
| Jul-2010 | 13 | 6 | 2 | 4 | 0 | 13 | 3 | 1 | 1 | 1 | 0 | 3 | 171 | 46 | 27 | 36 | 190 | 174 | 1 | 0 | 1 | 1 | 0 | 1 | 67319 | 21820 | 13502 | 13651 | 67664 | 67441 |
| Aug-2010 | 11 | 2 | 2 | 2 | 25 | 11 | 6 | 1 | 1 | 1 | 0 | 6 | 262 | 78 | 46 | 65 | 222 | 261 | 5 | 2 | 1 | 2 | 8 | 5 | 67035 | 21737 | 13451 | 13582 | 67409 | 67158 |
| Sep-2010 | 12 | 3 | 2 | 2 | 0 | 12 | 8 | 4 | 2 | 2 | 10 | 8 | 293 | 71 | 70 | 65 | 569 | 291 | 15 | 3 | 5 | 3 | 23 | 15 | 66707 | 21656 | 13374 | 13508 | 66807 | 66832 |
| Oct-2010 | 22 | 6 | 7 | 6 | 30 | 22 | 10 | 5 | 2 | 3 | 0 | 10 | 329 | 90 | 71 | 61 | 339 | 315 | 8 | 2 | 3 | 2 | 0 | 8 | 66338 | 21553 | 13292 | 13435 | 66438 | 66477 |
| Nov-2010 | 9 | 2 | 1 | 2 | 12 | 9 | 13 | 5 | 2 | 1 | 34 | 13 | 405 | 119 | 67 | 101 | 418 | 417 | 10 | 1 | 3 | 2 | 0 | 10 | 65901 | 21426 | 13220 | 13328 | 65974 | 66027 |
| Dec-2010 | 19 | 7 | 7 | 2 | 27 | 19 | 13 | 2 | 4 | 1 | 34 | 13 | 369 | 106 | 74 | 80 | 335 | 376 | 6 | 0 | 2 | 2 | 0 | 6 | 65494 | 21311 | 13141 | 13235 | 65578 | 65613 |
| Jan-2011 | 17 | 7 | 7 | 5 | 0 | 17 | 16 | 6 | 4 | 2 | 17 | 16 | 243 | 71 | 49 | 54 | 313 | 245 | 46 | 20 | 5 | 9 | 73 | 46 | 65172 | 21207 | 13076 | 13165 | 65175 | 65288 |
| Feb-2011 | 14 | 4 | 5 | 5 | 0 | 14 | 14 | 2 | 4 | 5 | 46 | 14 | 183 | 53 | 30 | 35 | 221 | 185 | 11 | 3 | 3 | 4 | 35 | 11 | 64950 | 21145 | 13032 | 13118 | 64873 | 65064 |
| Mar-2011 | 22 | 5 | 6 | 8 | 16 | 22 | 14 | 3 | 1 | 0 | 0 | 14 | 173 | 47 | 39 | 32 | 134 | 171 | 9 | 3 | 2 | 1 | 0 | 9 | 64732 | 21087 | 12984 | 13077 | 64723 | 64848 |
| Apr-2011 | 23 | 9 | 4 | 6 | 0 | 23 | 15 | 3 | 4 | 4 | 0 | 15 | 171 | 53 | 39 | 42 | 174 | 177 | 14 | 6 | 1 | 0 | 38 | 14 | 64509 | 21016 | 12935 | 13026 | 64511 | 64618 |
| May-2011 | 25 | 5 | 5 | 4 | 0 | 25 | 16 | 3 | 2 | 3 | 0 | 16 | 186 | 44 | 46 | 33 | 279 | 184 | 12 | 4 | 3 | 2 | 0 | 12 | 64270 | 20960 | 12880 | 12983 | 64232 | 64380 |
| Jun-2011 | 21 | 7 | 4 | 3 | 18 | 21 | 18 | 7 | 4 | 1 | 0 | 18 | 245 | 68 | 37 | 42 | 152 | 242 | 28 | 9 | 10 | 6 | 41 | 28 | 63958 | 20869 | 12833 | 12923 | 64021 | 64071 |
| Jul-2011 | 28 | 3 | 3 | 4 | 20 | 28 | 23 | 11 | 5 | 7 | 24 | 23 | 223 | 67 | 42 | 39 | 357 | 232 | 15 | 7 | 3 | 1 | 0 | 15 | 63669 | 20781 | 12779 | 12873 | 63620 | 63772 |
| Aug-2011 | 54 | 16 | 11 | 5 | 89 | 54 | 25 | 3 | 3 | 1 | 21 | 25 | 350 | 117 | 74 | 71 | 322 | 341 | 23 | 2 | 2 | 4 | 0 | 23 | 63217 | 20643 | 12695 | 12786 | 63188 | 63329 |
| Sep-2011 | 29 | 5 | 7 | 5 | 21 | 29 | 21 | 9 | 5 | 3 | 53 | 21 | 399 | 115 | 84 | 83 | 315 | 405 | 17 | 5 | 3 | 5 | 23 | 17 | 62751 | 20509 | 12598 | 12688 | 62776 | 62857 |
| Oct-2011 | 21 | 5 | 2 | 5 | 47 | 21 | 24 | 5 | 4 | 4 | 29 | 24 | 507 | 162 | 101 | 121 | 288 | 511 | 14 | 4 | 3 | 3 | 51 | 14 | 62185 | 20333 | 12485 | 12558 | 62361 | 62287 |
| Nov-2011 | 37 | 13 | 9 | 3 | 59 | 37 | 29 | 9 | 4 | 7 | 0 | 29 | 565 | 176 | 105 | 120 | 547 | 585 | 14 | 5 | 1 | 4 | 52 | 14 | 61540 | 20130 | 12366 | 12424 | 61703 | 61621 |
| Dec-2011 | 30 | 6 | 7 | 4 | 31 | 30 | 20 | 5 | 4 | 1 | 26 | 20 | 521 | 157 | 102 | 108 | 756 | 490 | 29 | 11 | 8 | 4 | 56 | 29 | 60940 | 19951 | 12255 | 12297 | 60834 | 61052 |
| Jan-2012 | 32 | 5 | 6 | 7 | 29 | 32 | 20 | 6 | 5 | 1 | 26 | 20 | 473 | 160 | 94 | 103 | 301 | 470 | 13 | 2 | 2 | 3 | 0 | 13 | 60402 | 19778 | 12150 | 12181 | 60478 | 60517 |
| Feb-2012 | 31 | 9 | 9 | 7 | 0 | 31 | 22 | 7 | 1 | 4 | 28 | 22 | 523 | 158 | 115 | 106 | 406 | 540 | 28 | 10 | 4 | 6 | 32 | 28 | 59798 | 19594 | 12018 | 12061 | 60012 | 59896 |
| Mar-2012 | 33 | 13 | 4 | 10 | 88 | 33 | 26 | 6 | 7 | 6 | 0 | 26 | 548 | 187 | 110 | 105 | 663 | 542 | 25 | 6 | 5 | 7 | 32 | 25 | 59166 | 19382 | 11885 | 11940 | 59229 | 59269 |
| Apr-2012 | 22 | 5 | 5 | 4 | 36 | 22 | 20 | 9 | 4 | 7 | 32 | 20 | 557 | 166 | 98 | 124 | 529 | 555 | 14 | 5 | 3 | 3 | 0 | 14 | 58553 | 19197 | 11773 | 11804 | 58632 | 58657 |
| May-2012 | 30 | 18 | 6 | 5 | 0 | 30 | 18 | 7 | 3 | 4 | 34 | 18 | 567 | 169 | 112 | 121 | 554 | 552 | 101 | 36 | 21 | 23 | 257 | 97 | 57837 | 18967 | 11629 | 11653 | 57787 | 57959 |
| Jun-2012 | 27 | 6 | 6 | 3 | 35 | 27 | 31 | 8 | 10 | 8 | 36 | 33 | 590 | 183 | 108 | 123 | 561 | 603 | 15 | 3 | 2 | 1 | 0 | 15 | 57174 | 18767 | 11509 | 11512 | 57155 | 57280 |
| Jul-2012 | 26 | 7 | 5 | 8 | 74 | 26 | 29 | 13 | 5 | 3 | 0 | 28 | 669 | 226 | 141 | 129 | 717 | 664 | 20 | 3 | 5 | 1 | 67 | 20 | 56430 | 18518 | 11356 | 11368 | 56297 | 56540 |
| Aug-2012 | 25 | 5 | 2 | 5 | 39 | 25 | 36 | 12 | 6 | 8 | 104 | 36 | 793 | 261 | 168 | 181 | 882 | 784 | 13 | 6 | 2 | 1 | 0 | 13 | 55563 | 18234 | 11174 | 11177 | 55272 | 55681 |
| Sep-2012 | 27 | 9 | 3 | 5 | 0 | 27 | 39 | 10 | 7 | 16 | 34 | 37 | 647 | 201 | 106 | 146 | 953 | 681 | 9 | 2 | 0 | 2 | 0 | 9 | 54841 | 18012 | 11045 | 11021 | 54285 | 54925 |
| Oct-2012 | 27 | 9 | 6 | 6 | 0 | 27 | 27 | 11 | 7 | 6 | 78 | 28 | 841 | 275 | 169 | 163 | 839 | 832 | 9 | 5 | 2 | 1 | 0 | 9 | 53937 | 17712 | 10863 | 10843 | 53368 | 54028 |
| Nov-2012 | 24 | 4 | 8 | 4 | 35 | 24 | 38 | 15 | 8 | 8 | 71 | 40 | 770 | 262 | 162 | 170 | 673 | 748 | 7 | 4 | 1 | 2 | 0 | 7 | 53098 | 17427 | 10687 | 10656 | 52589 | 53209 |
| Dec-2012 | 15 | 1 | 6 | 4 | 0 | 15 | 45 | 14 | 4 | 6 | 41 | 40 | 839 | 266 | 154 | 146 | 756 | 874 | 6 | 2 | 2 | 1 | 0 | 6 | 52193 | 17144 | 10522 | 10498 | 51792 | 52274 |

## Landmark Date – January 2013

| Date | REO | | | | | | Charge Off | | | | | | Payoff | | | | | | Others | | | | | | Survival | | | | | |
|---|---|---|---|---|---|---|---|---|---|---|---|---|---|---|---|---|---|---|---|---|---|---|---|---|---|---|---|---|---|---|
| | Original | Uniform | Vertical | Horizontal | Single | Backward | Original | Uniform | Vertical | Horizontal | Single | Backward | Original | Uniform | Vertical | Horizontal | Single | Backward | Original | Uniform | Vertical | Horizontal | Single | Backward | Original | Uniform | Vertical | Horizontal | Single | Backward |
| Jan-2013 | 0 | 0 | 0 | 0 | 0 | 0 | 0 | 0 | 0 | 0 | 0 | 0 | 0 | 0 | 0 | 0 | 0 | 0 | 0 | 0 | 0 | 0 | 0 | 0 | 608204 | 116879 | 121685 | 121475 | 599403 | 608007 |
| Feb-2013 | 83 | 23 | 15 | 13 | 50 | 83 | 143 | 32 | 26 | 18 | 168 | 146 | 3350 | 635 | 665 | 640 | 3454 | 3330 | 34 | 5 | 8 | 5 | 31 | 34 | 604594 | 116184 | 120984 | 120786 | 595700 | 604414 |
| Mar-2013 | 97 | 13 | 17 | 25 | 140 | 97 | 159 | 28 | 41 | 36 | 154 | 163 | 3850 | 727 | 757 | 731 | 3623 | 3804 | 37 | 8 | 7 | 11 | 5 | 37 | 600450 | 115408 | 120154 | 119990 | 591778 | 600302 |
| Apr-2013 | 103 | 21 | 20 | 29 | 107 | 102 | 177 | 32 | 35 | 42 | 233 | 177 | 4337 | 835 | 871 | 860 | 4020 | 4417 | 34 | 7 | 8 | 6 | 35 | 34 | 595799 | 114513 | 119206 | 119067 | 587383 | 595571 |
| May-2013 | 75 | 18 | 14 | 14 | 79 | 75 | 194 | 39 | 39 | 38 | 202 | 192 | 4984 | 888 | 1006 | 996 | 5101 | 4946 | 16 | 4 | 4 | 3 | 14 | 16 | 590530 | 113564 | 118145 | 118014 | 581987 | 590341 |
| Jun-2013 | 104 | 15 | 20 | 26 | 66 | 102 | 191 | 38 | 40 | 31 | 77 | 187 | 4913 | 934 | 994 | 4673 | 4893 | 29 | 9 | 10 | 8 | 16 | 29 | 585293 | 112568 | 117086 | 116978 | 577155 | 585128 |
| Jul-2013 | 124 | 22 | 16 | 28 | 74 | 126 | 210 | 34 | 43 | 36 | 210 | 211 | 5239 | 984 | 1099 | 1041 | 5039 | 5429 | 27 | 7 | 6 | 6 | 7 | 27 | 579693 | 111521 | 115917 | 115872 | 571825 | 579334 |
| Aug-2013 | 136 | 28 | 23 | 30 | 134 | 126 | 165 | 31 | 37 | 32 | 155 | 165 | 4063 | 741 | 826 | 843 | 4245 | 4012 | 17 | 4 | 3 | 4 | 19 | 17 | 575312 | 110717 | 115025 | 114966 | 567272 | 575012 |
| Sep-2013 | 113 | 19 | 21 | 23 | 109 | 113 | 152 | 28 | 30 | 26 | 76 | 150 | 3051 | 592 | 601 | 591 | 3423 | 3068 | 22 | 3 | 5 | 5 | 38 | 22 | 571974 | 110075 | 114370 | 114319 | 563626 | 571659 |
| Oct-2013 | 136 | 32 | 30 | 29 | 159 | 128 | 159 | 34 | 27 | 32 | 83 | 160 | 2991 | 539 | 590 | 592 | 2916 | 2956 | 20 | 2 | 8 | 6 | 16 | 20 | 568668 | 109468 | 113713 | 113662 | 560452 | 568394 |
| Nov-2013 | 125 | 26 | 33 | 25 | 11 | 124 | 121 | 17 | 26 | 25 | 141 | 120 | 2777 | 510 | 561 | 560 | 2857 | 2826 | 19 | 6 | 7 | 3 | 0 | 19 | 565626 | 108909 | 113099 | 113036 | 557443 | 565304 |
| Dec-2013 | 134 | 23 | 30 | 28 | 228 | 130 | 148 | 27 | 40 | 28 | 222 | 148 | 2970 | 528 | 618 | 566 | 2944 | 2962 | 15 | 1 | 4 | 3 | 15 | 15 | 562359 | 108330 | 112422 | 112366 | 554067 | 562048 |
| Jan-2014 | 130 | 31 | 26 | 28 | 42 | 130 | 102 | 23 | 28 | 23 | 266 | 100 | 2113 | 410 | 421 | 405 | 1978 | 2116 | 39 | 6 | 9 | 9 | 43 | 39 | 559975 | 107860 | 111941 | 111928 | 551705 | 559663 |
| Feb-2014 | 105 | 14 | 20 | 29 | 80 | 103 | 87 | 16 | 16 | 12 | 79 | 85 | 1994 | 358 | 417 | 377 | 2460 | 2003 | 47 | 9 | 11 | 13 | 89 | 47 | 557742 | 107463 | 111470 | 111504 | 548997 | 557423 |
| Mar-2014 | 119 | 22 | 24 | 20 | 126 | 121 | 99 | 25 | 21 | 24 | 119 | 98 | 2708 | 521 | 552 | 533 | 2775 | 2706 | 191 | 45 | 42 | 38 | 248 | 186 | 554625 | 106850 | 110836 | 110884 | 545729 | 554311 |
| Apr-2014 | 92 | 20 | 24 | 16 | 23 | 93 | 109 | 21 | 17 | 18 | 37 | 111 | 3149 | 585 | 658 | 595 | 2933 | 3174 | 51 | 9 | 10 | 8 | 19 | 51 | 551224 | 106215 | 110136 | 110238 | 542717 | 550881 |
| May-2014 | 95 | 18 | 20 | 18 | 225 | 95 | 109 | 15 | 18 | 21 | 163 | 113 | 3487 | 648 | 716 | 697 | 3441 | 3512 | 47 | 10 | 5 | 9 | 145 | 47 | 547486 | 105524 | 109372 | 109498 | 538743 | 547113 |
| Jun-2014 | 73 | 11 | 17 | 10 | 137 | 73 | 93 | 16 | 15 | 18 | 26 | 92 | 4076 | 762 | 819 | 810 | 3750 | 4175 | 51 | 8 | 14 | 11 | 38 | 51 | 543193 | 104727 | 108514 | 108642 | 534792 | 542722 |
| Jul-2014 | 65 | 6 | 11 | 12 | 37 | 65 | 106 | 20 | 18 | 17 | 56 | 108 | 4577 | 815 | 888 | 850 | 5169 | 4650 | 62 | 13 | 13 | 14 | 26 | 62 | 538383 | 103873 | 107583 | 107750 | 529504 | 537836 |
| Aug-2014 | 88 | 15 | 18 | 16 | 187 | 90 | 94 | 15 | 17 | 17 | 20 | 67 | 94 | 4151 | 740 | 842 | 852 | 3948 | 4292 | 63 | 12 | 14 | 10 | 0 | 63 | 533987 | 103091 | 106695 | 106849 | 525302 | 533296 |
| Sep-2014 | 75 | 17 | 15 | 17 | 125 | 75 | 95 | 17 | 18 | 16 | 88 | 101 | 4041 | 761 | 810 | 787 | 4300 | 4052 | 62 | 10 | 12 | 10 | 29 | 62 | 529714 | 102286 | 105842 | 106017 | 520760 | 529005 |
| Oct-2014 | 94 | 14 | 18 | 18 | 249 | 102 | 107 | 23 | 23 | 33 | 64 | 104 | 4148 | 794 | 857 | 846 | 4327 | 4272 | 73 | 18 | 13 | 12 | 131 | 73 | 525287 | 101437 | 104922 | 105117 | 515989 | 524453 |
| Nov-2014 | 68 | 11 | 20 | 14 | 62 | 68 | 87 | 15 | 18 | 17 | 76 | 85 | 3573 | 677 | 696 | 714 | 3943 | 3552 | 53 | 14 | 7 | 10 | 57 | 53 | 521506 | 100720 | 104185 | 104358 | 511851 | 520694 |
| Dec-2014 | 114 | 21 | 26 | 26 | 65 | 116 | 103 | 20 | 24 | 16 | 75 | 103 | 4201 | 800 | 835 | 843 | 3789 | 4350 | 62 | 17 | 11 | 9 | 88 | 62 | 517026 | 99862 | 103299 | 103454 | 507834 | 516062 |
| Jan-2015 | 95 | 17 | 17 | 20 | 135 | 95 | 73 | 11 | 14 | 14 | 71 | 73 | 3274 | 647 | 654 | 636 | 3427 | 3180 | 66 | 15 | 13 | 12 | 50 | 66 | 513518 | 99172 | 102599 | 102774 | 504151 | 512648 |
| Feb-2015 | 76 | 12 | 18 | 6 | 113 | 73 | 92 | 14 | 18 | 18 | 109 | 91 | 3962 | 708 | 810 | 773 | 3860 | 3950 | 70 | 14 | 13 | 14 | 107 | 70 | 509318 | 98424 | 101751 | 101952 | 499962 | 508463 |
| Mar-2015 | 93 | 18 | 23 | 17 | 223 | 94 | 91 | 25 | 13 | 17 | 69 | 96 | 5816 | 1038 | 1185 | 1118 | 5920 | 5680 | 73 | 16 | 13 | 18 | 96 | 72 | 503245 | 97327 | 100514 | 100785 | 493654 | 502520 |
| Apr-2015 | 85 | 19 | 17 | 22 | 62 | 83 | 95 | 22 | 27 | 14 | 37 | 93 | 5516 | 1010 | 1104 | 1034 | 5006 | 5520 | 92 | 20 | 13 | 14 | 205 | 94 | 497457 | 96256 | 99360 | 99694 | 488344 | 496729 |
| May-2015 | 69 | 14 | 14 | 17 | 57 | 69 | 81 | 9 | 11 | 16 | 0 | 80 | 5623 | 1005 | 1108 | 1144 | 4423 | 5520 | 229 | 52 | 56 | 48 | 428 | 232 | 491455 | 95176 | 98171 | 98469 | 483436 | 490827 |
| Jun-2015 | 76 | 12 | 16 | 11 | 111 | 76 | 96 | 24 | 16 | 16 | 36 | 97 | 6195 | 1178 | 1171 | 1209 | 5867 | 6230 | 71 | 15 | 15 | 8 | 0 | 71 | 485017 | 93947 | 96965 | 97210 | 477422 | 484353 |
| Jul-2015 | 91 | 18 | 17 | 15 | 181 | 92 | 92 | 19 | 20 | 27 | 129 | 91 | 5750 | 1035 | 1175 | 1153 | 5673 | 5950 | 96 | 21 | 24 | 16 | 90 | 97 | 478988 | 92854 | 95732 | 95996 | 471349 | 478121 |
| Aug-2015 | 77 | 21 | 10 | 14 | 111 | 74 | 82 | 20 | 16 | 15 | 103 | 82 | 5130 | 944 | 1035 | 982 | 5965 | 4875 | 74 | 16 | 17 | 18 | 37 | 71 | 473625 | 91853 | 94650 | 94971 | 465133 | 473019 |
| Sep-2015 | 77 | 13 | 15 | 18 | 36 | 77 | 73 | 12 | 12 | 8 | 76 | 73 | 4956 | 964 | 955 | 988 | 5335 | 5225 | 111 | 16 | 22 | 19 | 58 | 110 | 468408 | 90848 | 93650 | 93934 | 459628 | 467534 |
| Oct-2015 | 74 | 11 | 14 | 11 | 0 | 76 | 75 | 14 | 18 | 15 | 0 | 76 | 4737 | 881 | 979 | 918 | 4324 | 4606 | 131 | 23 | 21 | 28 | 114 | 134 | 463391 | 89919 | 92617 | 92963 | 455190 | 462639 |
| Nov-2015 | 68 | 9 | 14 | 8 | 0 | 68 | 45 | 10 | 15 | 5 | 148 | 45 | 3850 | 684 | 809 | 761 | 3539 | 3857 | 136 | 31 | 25 | 24 | 148 | 141 | 459292 | 89185 | 91771 | 92148 | 451355 | 458528 |
| Dec-2015 | 61 | 16 | 6 | 9 | 0 | 61 | 83 | 22 | 17 | 13 | 0 | 82 | 4659 | 828 | 920 | 956 | 4715 | 4573 | 87 | 14 | 17 | 14 | 68 | 85 | 454402 | 88305 | 90815 | 91152 | 446572 | 453725 |

## Appendix D – Figure of Hazard Curves

### Landmark Date – January 2010

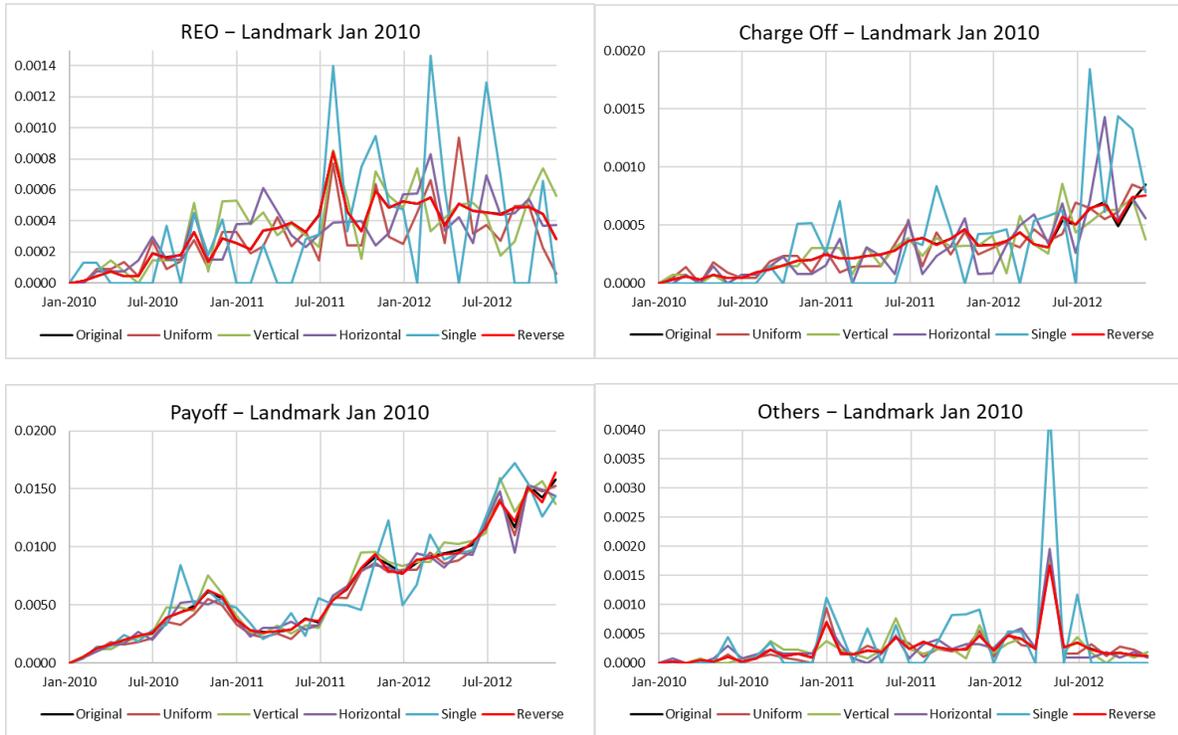

### Landmark Date – January 2013

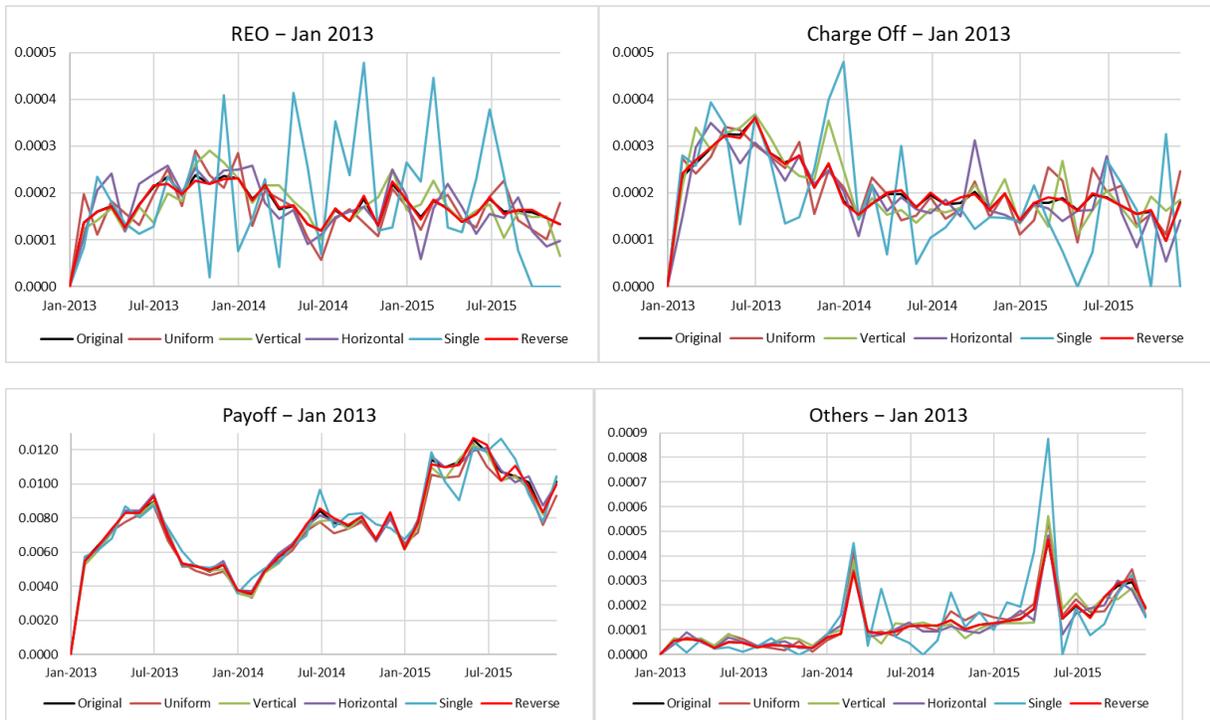

Appendix E

**Monthly Hazard of The Unstacked Original Dataset and The Super Dataset**

Let $n_t$ denote the number of individuals censored at time $t$. Time $t$ may be the censoring time due to other events or the end of the study. For example, $n_{10}$ denotes that there are $n_{10}$ individuals (not observations) with survival time equal to 10.

Let $d_t$ denote the number of individuals uncensored at time $t$. Time $t$ is the time to an event. For example, $d_5$ denotes that there are $d_5$ individuals (not observations) with event time equal to 5.

From month 1 ($m=1$) to month $M$, Table A1 shows the number of events, survivals, total observations and the monthly hazard.

| $m$ | # event | # observations of event individuals | # observations of survival individuals | $h_i$ (hazard) |
|---|---|---|---|---|
| 1 | $d_1$ | $d_1+d_2+...+d_M$ | $n_1+n_2+...+n_M$ | $d_1 / ((d_1+d_2+...+d_M)+(n_1+n_2+...+n_M))$ |
| 2 | $d_2$ | $d_2+d_3+...+d_M$ | $n_2+n_3+...+n_M$ | $d_2 / ((d_2+d_3+...+d_M)+(n_2+n_3+...+n_M))$ |
| 3 | $d_3$ | $d_3+d_4+...+d_M$ | $n_3+n_4+...+n_M$ | $d_3 / ((d_3+d_4+...+d_M)+(n_3+n_4+...+n_M))$ |
| ⋮ | ⋮ | ⋮ | ⋮ | ⋮ |
| $M$ | $d_M$ | $d_M$ | $n_M$ | $d_M / (d_M+n_M)$ |

Table A1: Monthly hazard of the unstacked original dataset

Note that "# number of the observations of event individuals" is the number of observations of individuals experiencing the event. When individuals are censored ($n$) or uncensored ($d$) at time 3, say, their observations are in $m=1, 2$, and 3. Thus, there are $n_3$ observations, or $d3$ observations in $m=1, 2$ and 3.

For super dataset, the number of events, survivals, total observations and the monthly hazard are in Table A2.

| $m$ | # event | # observations of event individuals | # observations of survival individuals | $h_i$ (hazard) |
|---|---|---|---|---|
| 1 | $1d_1$ | $1(d_1+d_2+...+d_M)$ | $1(n_1+n_2+...+n_M)$ | $d_1 / ((d_1+d_2+...+d_M)+(n_1+n_2+...+n_M))$ |
| 2 | $2d_2$ | $2(d_2+d_3+...+d_M)$ | $2(n_2+n_3+...+n_M)$ | $d_2 / ((d_2+d_3+...+d_M)+(n_2+n_3+...+n_M))$ |
| 3 | $3d_3$ | $3(d_3+d_4+...+d_M)$ | $3(n_3+n_4+...+n_M)$ | $d_3 / ((d_3+d_4+...+d_M)+(n_3+n_4+...+n_M))$ |
| ⋮ | ⋮ | ⋮ | ⋮ | ⋮ |
| $M$ | $Md_M$ | $Md_M$ | $Mn_M$ | $d_M / (d_M+n_M)$ |

Table A2: Monthly hazard of the super dataset

Refer to the illustration in Section II,

In the unstacked original dataset, total # of survival observations = $n_1+n_2+...+n_M$.

In the super dataset, total # of survival observations = $1n_1+2n_2+...+Mn_M = \sum_{m=1}^{M} mn_n$.

In the unstacked original dataset, total # of event observations = $d_1+d_2+...+d_M$.

In the super dataset, total # of event observations = $1d_1+2d_2+...+Md_M = \sum_{m=1}^{M} md_n$.

Therefore, the monthly hazard in the super dataset is identical to the monthly hazard in the unstacked original dataset.

**Monthly Hazard of The Vertical Random Landmark Sample**

The following derivation is to obtain the expected number of events, non-events, and, then, the expected monthly hazard of the vertical sample with all the possible landmarks being selected.

Let an individual's survival time be $n$ ($t=1 \ldots n$). The following table shows the event index of the original dataset and all the stacks from each landmark arranged side-by-side by the same time $t$.

| Survival time | Original Data | Row Count | Super Dataset | | | | | Row Count |
|---|---|---|---|---|---|---|---|---|
| | | | 1st stack | 2nd stack | 3rd stack | … | $n$th stack | |
| 1 | x | 1 | x | | | … | | 1 |
| 2 | x | 1 | x | x | | | | 2 |
| 3 | x | 1 | x | x | x | | | 3 |
| ⋮ | ⋮ | 1 | ⋮ | ⋮ | ⋮ | | | ⋮ |
| $n$ | x | 1 | x | x | x | | x | $n$ |

Table A3: The Original Dataset and The Super Dataset

Let the 1st stack refer to the subsequent observations selected after the first landmark including the first landmark observation, let the 2nd stack refer to the subsequent observations selected after the second landmark including the second landmark observation, and so forth. Thus, the number of observations of the 1st stack is $n$, the 2nd stack has $n-1$ observations, and, the $n$th stack (last stack) has 1 observation.

When selecting a random sample of landmarks from these $n$ landmarks in the original dataset, each landmark can either be selected or unselected. Thus, there are $2^n$ combinations of stacks can be selected. The number of stacks selected into a random sample of landmarks ranges from 0 to $n$, and the number can be denoted as $_nC_i$ that is defined as $\binom{n}{i} = \frac{n!}{i!(n-i)!}$, $i=0,\ldots,n$ where $i$ denote the number of landmarks being selected. Hence, when $i=0$, it means no landmark is selected into the random sample, and, the random sample is simply an empty set. When $i=5$, say, there are $_nC_5$ combinations of landmarks selected into the random sample of 5 landmarks. The vertical sample then has the 5 selected landmarks and their subsequent observations.

After arranging these $2^n$ combinations side-by-side by $t$, the number of observations (row count) of the first row is positive if the first stack is selected, the number of observations (row count) of the second row is positive if the first stack or the second stack is selected, and so forth.

If an individual has $n$ landmarks and when the first stack is selected (only one choice= $_1C_1$), there are $2^{n-1}$ combinations of the remaining $n-1$ stacks to be selected. These remaining $n-1$ stacks are the 2nd stack, the 3rd stack, …, the $n$th stack. Thus, the sum of the row counts at $t=1$ from all the combinations of stacks selected is $_1C_1 \times 2^{n-1}$.

The 1st stack and the 2nd stack contribute to the row count of $t=2$. When either one of the first or the second stack, or both is selected, the number of stacks being selected can be expressed by the coefficients of the expansion of $(a+b)^2$. The expansion is ${}_2C_0 a^2 b^0 + {}_2C_1 a^1 b^1 + {}_2C_2 a^0 b^2$.

The first term ${}_2C_0 a^2 b^0$ can be regarded as none of the first and the second stack is selected. Thus, the row count from this term is $0 \cdot {}_2C_0 = 0$.

The second term ${}_2C_1 a^1 b^1$ can be regarded as either the first or the second stack is selected. Thus, the row count from this term is $1 \cdot {}_2C_1 = 2$.

The third term ${}_2C_2 a^0 b^2$ can be regarded as both the first and the second stack are selected. Thus, row count from this term is $2 \cdot {}_2C_2 = 2$.

After excluding the first and the second stack, the remaining stacks are from $t=3$ to $n$ with $2^{n-2}$ combinations. Thus, the sum of row counts at $t=2$ from all the combinations of stacks selected is $(0 \cdot {}_2C_0 + 1 \cdot {}_2C_1 + 2 \cdot {}_2C_2) \cdot 2^{n-2}$.

In general, the sum of row counts at $t=i$ ($i=1$ to $n$) from all the combinations of stacks selected can be expressed by the expansion of $(a+b)^i$, and it is

$$0 \cdot {}_iC_0 + 1 \cdot {}_iC_1 + 2 \cdot {}_iC_2 + \ldots + i \cdot {}_iC_i = \sum_{j=0}^{i} j \binom{n}{j} = \sum_{j=0}^{i} \frac{i!}{(j-1)!(i-j)!} = i \sum_{j-1=0}^{i-1} \binom{i-1}{j-1} = i \cdot 2^{i-1}.$$

(Note that $\binom{n}{0} + \binom{n}{1} + \cdots + \binom{n}{n} = \sum_{i=0}^{n} \binom{n}{i} = 2^n$.)

Since there are $2^i$ combinations of number of landmarks being selected, the expected value of the $i$th row count is $\frac{i \cdot 2^{i-1}}{2^i} = \frac{i}{2}$.

Therefore, the expected value of the sum of row counts at $t$ ($t=1, \ldots, n$) of this individual with $n$ landmarks is $\frac{t}{2}$, $t=1, \ldots, n$.

If the individual experiences an event, the expected value of the last row count becomes the expected value of the number of events.

For the vertical landmark sample, the expected value of the number of events, event observations, survival observations, and the hazard are listed in the following table.

| m | Expected Value of Number of Events | Expected Value of Number of Observations of Event Individuals | Expected Value of Number of Observations of Survival Individuals | Expected $h_m$ (Hazard) |
|---|---|---|---|---|
| 1 | $d_1 \cdot \frac{1}{2}$ | $d_1 \cdot \frac{1}{2} + d_2 \cdot \frac{1}{2} + \ldots + d_M \cdot \frac{1}{2}$ | $n_1 \cdot \frac{1}{2} + n_2 \cdot \frac{1}{2} + \ldots + n_M \cdot \frac{1}{2}$ | $d_1 / ((d_1 + d_2 + \ldots + d_M) + (n_1 + n_2 + \ldots + n_M))$ |
| 2 | $d_2 \cdot \frac{2}{2}$ | $d_2 \cdot \frac{2}{2} + d_3 \cdot \frac{2}{2} + \ldots + d_M \cdot \frac{2}{2}$ | $n_2 \cdot \frac{2}{2} + n_3 \cdot \frac{2}{2} + \ldots + n_M \cdot \frac{2}{2}$ | $d_2 / ((d_2 + d_3 + \ldots + d_M) + (n_2 + n_3 + \ldots + n_M))$ |
| 3 | $d_3 \cdot \frac{3}{2}$ | $d_3 \cdot \frac{3}{2} + d_4 \cdot \frac{3}{2} + \ldots + d_M \cdot \frac{3}{2}$ | $n_3 \cdot \frac{3}{2} + n_4 \cdot \frac{3}{2} + \ldots + n_M \cdot \frac{3}{2}$ | $d_3 / ((d_3 + d_4 + \ldots + d_M) + (n_3 + n_4 + \ldots + n_M))$ |
| ⋮ | ⋮ | ⋮ | ⋮ | ⋮ |
| M | $d_M \cdot \frac{M}{2}$ | $d_M \cdot \frac{M}{2}$ | $n_M \cdot \frac{M}{2}$ | $d_M / (d_M + n_M)$ |

Table A4: The expected value of the vertical landmark sample

As we can see, the expected monthly hazard of the random landmark sample is identical to of the original dataset and the super dataset. Therefore, the monthly hazard of the vertical landmark sampling is unbiased estimator of the unstacked original dataset.

**Row Counts of The Vertical Landmark Sample with a Predetermined Sample Size**

Suppose an individual has $M$ months of history (survival time equal to $M$). As it has been shown that the row count in $m$th month in the super dataset is $m$, $m=1$ to $M$. Therefore, when randomly selecting $n$ landmarks from the total of $M$ landmarks ($n<M$), the maximum of the $m$th row count in a random landmark sample cannot exceed $m$.

In a random landmark sample after arranging all the stacks side-by-side by $m$, in order for the row count in the 1st row to be 1, the 1st landmark must be selected, and there are combinations of $1\binom{1}{1}\binom{M-1}{n-1}$. That is, the 1st landmark is selected with one count denoted as $1\binom{1}{1}$, and $n-1$ landmarks are selected from the remaining $M-1$ landmarks denoted as $\binom{M-1}{n-1}$. In other words, the sum of the 1st month row count from all the possible combinations of $n$ landmarks is $1\binom{1}{1}\binom{M-1}{n-1}$. If the 1st landmark is not selected, then the 1st month row count is zero.

Similarly, in order for the row count in the 2nd month row to be positive (1 or 2), one of the 1st and the 2nd landmark or both must be selected, and the sum of the row counts in the 2nd month row from all possible combinations of $n$ landmarks is $1\binom{2}{1}\binom{M-2}{n-1}+2\binom{2}{2}\binom{M-2}{n-2}$. That is, one of the 1st and 2nd landmark is selected with one count denoted as $1\binom{2}{1}$, and $n-1$ landmarks are selected from the remaining $M-2$ landmarks denoted as $\binom{M-2}{n-1}$. When both 1st and 2nd landmark are selected, the row count is $2\binom{2}{2}$, and $n-2$ landmarks are selected from the remaining $M-2$ landmarks denoted as $\binom{M-2}{n-2}$.

Consequently, in order for the row count in the $i$th row to be positive, at least one of landmarks from 1st to $i$th landmark must be selected. Therefore, when randomly selecting $n$ landmarks ($n<M$) from the total $M$ landmarks, the sum of row counts in $i$th month from all possible combinations is

$$1\binom{m}{1}\binom{M-m}{n-1}+2\binom{m}{2}\binom{M-m}{n-2}+3\binom{m}{3}\binom{M-m}{n-3}+\ldots$$

The expression for the above sum of row counts can be formulized as

$$\sum_{i=1}^{m} i\binom{m}{i}\binom{M-m}{n-i}, m=1 \text{ to } M, i \leq m \leq M.$$

The expression is equal to 0 when $M-m < n-i$.

Then, the expression can be simplified using the Vandermonde Convolution formula in Combinatorial Methods with Computer Applications by Gross (2008, CRC Press)

$$\sum_{i=1}^{m} i\binom{m}{i}\binom{M-m}{n-i}=m\sum_{i=1}^{m}\binom{m-1}{i-1}\binom{M-m}{n-i}=m\binom{M-1}{n-1}$$

When selecting a sample of $n$ landmarks from the total of $M$ landmarks, there are $\binom{M}{n}$ combinations.

Thus, the expected the row count in $m$th row of the individual is $\frac{m\binom{M-1}{n-1}}{\binom{M}{n}} = m\frac{n}{M} = mp$ where $p$ is the sampling rate.

Consequently, for a random landmark sample with sampling rate $p$, the expected value of the number of events, event observations, survival observations, and the hazard are listed in the following table.

| $m$ | Expected Value of Number of Events | Expected Value of Number of Observations of Event Individuals | Expected Value of Number of Observations of Survival Individuals | Expected $h_m$ (Hazard) |
|---|---|---|---|---|
| 1 | $d_1 \cdot 1p$ | $d_1 \cdot 1p + d_2 \cdot 1p + \ldots + d_M \cdot 1p$ | $n_1 \cdot 1p + n_2 \cdot 1p + \ldots + n_M \cdot 1p$ | $d_1 / ((d_1+d_2+\ldots+d_M)+(n_1+n_2+\ldots+n_M))$ |
| 2 | $d_2 \cdot 2p$ | $d_2 \cdot 2p + d_3 \cdot 2p + \ldots + d_M \cdot 2p$ | $n_2 \cdot 2p + n_3 \cdot 2p + \ldots + n_M \cdot 2p$ | $d_2 / ((d_2+d_3+\ldots+d_M)+(n_2+n_3+\ldots+n_M))$ |
| 3 | $d_3 \cdot 3p$ | $d_3 \cdot 3p + d_4 \cdot 3p + \ldots + d_M \cdot 3p$ | $n_3 \cdot 3p + n_4 \cdot 3p + \ldots + n_M \cdot 3p$ | $d_3 / ((d_3+d_4+\ldots+d_M)+(n_3+n_4+\ldots+n_M))$ |
| ⋮ | ⋮ | ⋮ | ⋮ | ⋮ |
| M | $d_M \cdot Mp$ | $d_M \cdot Mp$ | $n_M \cdot Mp$ | $d_M / (d_M+n_M)$ |

Table A5: The expected value of the random landmark sample with a sampling rate

As we can see, the expected monthly hazard of the random landmark sample with sampling rate $p$ is identical to of the original dataset, the super dataset, and the vertical sample. Therefore, monthly hazard of the random landmark sampling with sampling rate $p$ is unbiased estimator of the unstacked original dataset. Moreover, when the sampling rate $p$ is 1, it becomes the super dataset.

**The Unbiased Estimator of The Horizontal Landmark Sampling**

Since a random sample is created from each landmark – the stratum, the monthly hazard is unbiased estimator of the original dataset.

**The Unbiased Estimator of The Uniformly Spaced Landmark Sampling**

After deleting the stacks with equidistance from the super data, it becomes the uniform sample. It is equivalent to deleting stacks from the super dataset, and those stacks are from the landmarks fall into the "equidistance". Similar to the super dataset, therefore, the number of events and survivals are also proportional to of the unstacked original dataset. The monthly hazard, hence, is an unbiased estimator of the unstacked original dataset.

**The Unbiased Estimator of The Backward Landmark Sample**

Table A1 can be regarded as a specific landmark from the population. Supposing the sampling rate is $\rho$, in $m$th month, there are $\rho d_m$ events. The survival observations in $m$th month are the preceding observations from event individuals after $m$th month, and the non-event individuals in and after $m$th month.

Then, the number of survival observations in $m$th month are $\rho(d_{m+1}+ \ldots +d_M) + \rho(n_m + n_{m+1} + \ldots + n_M)$. Therefore, the expectation of the hazard in $m$th month is

$$E[\rho d_m/(\rho d_m + \rho(d_{m+1} + \ldots + d_M) + \rho(n_m + n_{m+1} + \ldots + n_M))] = \frac{d_m}{(d_m + \cdots + d_M) + (n_m + \cdots + n_M)},$$

an unbiased estimator of the unstacked original dataset.

## The Variance of The Unstacked Original Dataset

Recall that $d_i$ is the number of individuals with survival time $i$ and the event occurs at time $i$. That is, the value of the event index for these individuals become 1 at time $i$ and 0 otherwise.

The variance of the $i$th monthly hazard in the population is equal to

$$\frac{d_i(1-\mu_i)^2 + ((d_{i+1} + \cdots + d_M) + (n_i + \cdots + n_M))(0-\mu_i)^2}{(d_i + \cdots + d_M) + (n_i + \cdots + n_M)}$$

$$= \frac{d_i(1-\mu_i)^2 + ((d_i + \cdots + d_M) + (n_i + \cdots + n_M))(0-\mu_i)^2 - d_i(0-\mu_i)^2}{(d_i + \cdots + d_M) + (n_i + \cdots + n_M)}$$

$$= \frac{d_i(1-2\mu_i) + \frac{d_i}{\mu_i}\mu_i^2}{\frac{d_i}{\mu_i}}$$

$$= \mu_i(1-\mu_i)$$

where $\mu_i = \frac{d_i}{(d_i + \cdots + d_M) + (n_i + \cdots + n_M)}$.

Since the monthly number of observations of event or non-event in the super dataset is $m$ (the number of month) multiplying the number of observations in the population, the variance of the $i$th monthly hazard in the super dataset is identical to of the original dataset. (Simply multiply both the numerator and the denominator by $m$ in the above equations.)

## The Standard Error of the Monthly Hazard of The Vertical Landmark Sample

Suppose a vertical landmark sample of $\rho\%$ is selected and the variance of the $i$th month hazard is $Var(X)$, then, the variance of the means of the $i$th month hazards from all samples can be approximated by $\left(1 - \frac{n}{N}\right)\frac{Var(X)}{n}$ where $n$ is the sample size (total number of observations) in $i$th month, $N$ is the size of the super dataset in $i$th month.

The expression $\left(1 - \frac{n}{N}\right)$ is the finite population correction factor (Johnson, 2000). Note that $\frac{n}{N} = \rho\%$, the sampling rate. Also note that $Var(X)$ should be the variance of the population but it is often not available so it is approximated by the sample variance. Therefore, when selecting a $\rho\%$ vertical landmark sample, the standard error of $i$th month hazard can be approximated by $(1 - \rho\%)\frac{Var(X)}{n}$.